\documentclass[letterpaper, 10 pt, conference]{ieeeconf}  

\IEEEoverridecommandlockouts                              

\usepackage{cite}
\usepackage{amsmath,amssymb,amsfonts}
\usepackage{algorithmic}
\usepackage{graphicx}
\usepackage{textcomp}
\usepackage{xcolor}
\usepackage{todonotes}
\usepackage{cite}
\usepackage{url}
\def\BibTeX{{\rm B\kern-.05em{\sc i\kern-.025em b}\kern-.08em
    T\kern-.1667em\lower.7ex\hbox{E}\kern-.125emX}}
\begin{document}

\title{\LARGE \bf The eyes and hearts of UAV pilots: observations of  physiological responses in real-life scenarios

}

\author{Alexandre Duval$^{1}$, Anita Paas$^{2}$, Abdalwhab Abdalwhab$^{1}$ and David St-Onge$^{1}$
\thanks{*We thank NSERC USRA and Discovery programs for their financial support. We also acknowledge the support provided by Calcul Québec and Compute Canada.}
\thanks{$^{1}$Alexandre Duval, Abdalwhab Abdalwhab and David St-Onge are with the Lab INIT Robots, Department of Mechanical Engineering,
        Ecole de technologie supérieure, Canada
        {\tt\small name.surname@etsmtl.ca}}%
\thanks{$^{2}$Anita Paas is with the Department of Psychology, Concordia University,
        Canada
        {\tt\small anita.paas@concordia.ca}}%
}

\maketitle
\thispagestyle{empty}
\pagestyle{empty}

\begin{abstract}
The drone industry is diversifying and the number of pilots increases rapidly. In this context, flight schools need adapted tools to train pilots, most importantly with regard to their own awareness of their physiological and cognitive limits. In civil and military aviation, pilots can train themselves on realistic simulators to tune their reaction and reflexes, but also to gather data on their piloting behavior and physiological states. It helps them to improve their performances. Opposed to cockpit scenarios, drone teleoperation is conducted outdoor in the field, thus with only limited potential from desktop simulation training. This work aims to provide a solution to gather pilots behavior out in the field and help them increase their performance. We combined advance object detection from a frontal camera to gaze and heart-rate variability measurements. We observed pilots and analyze their behavior over three flight challenges. We believe this tool can support pilots both in their training and in their regular flight tasks.
\end{abstract}

\section{Introduction}
The industry of teleoperated drones for service, such as in infrastructure inspection, crops monitoring and cinematography, has expanded at least as fast as the technology that supports it over the past decade. However, in most countries the regulation is only slowly adapting. Nevertheless, several regulating bodies already recognized human factors as a core contributor to flight hazards. While core technical features of the aerial systems are evolving, namely autonomy, flight performances and onboard sensing, the human factors of UAV piloting stay mostly uncharted territory.


Physiological measures, including eye-based measures (changes in pupil diameter, gaze-based data, and blink rate), heart rate variability, and skin conductance, are valuable indirect measures of cognitive workload. These measures are increasingly used to measure workload level during a task and are being integrated into interfaces and training applications to optimize performance and training programs \cite{Liu2016}.

Eye-tracking glasses can be used to monitor training progress during various levels of task load. In a task requiring operators to track targets and other vehicles on a map, Coyne and Sibley \cite{Coyne2015} found a significant decrease in operator situational awareness when task load was high, which was related to reduced eye gaze spent on the map. This suggests that eye gaze may be useful as a predictor of situational awareness. Further, Memar and Esfahani \cite{Memar2018} found that gaze-based data were related to target detection and situational awareness in a tele-exploration task with a swarm of robots.


Thus, multisensory configurations can be more robust to capture cognitive load. While each sensor is susceptible to some noise, these sources of noise do not overlap between sensors, such as HRV not influenced by luminance.
This works aims at extracting gaze behavior and so we also gather pupil diameter for cognitive load estimation. However, we added another device extract HRV metrics and enhance our cognitive load estimation.

Section~\ref{sec:litrev} opens the path with an overview of the various inspirational domains to this work. We then build our solution on a biophysical capturing software (sec.~\ref{sec:biop}) and a detector trained on a custom dataset (sec.~\ref{sec:learn}). Finally, we present the results of a small user study in sec.~\ref{sec:exp} and discuss our observations of the pilots behaviors.

\section{Related Works}\label{sec:litrev}
\subsection{On gaze-based behavioral studies}
The benefit of eye tracking is that we can measure gaze behavior in addition to changes in pupil diameter. Gaze behavior can provide information about the most efficient way to scan and monitor multiple sources of input. For example, in surveillance tasks, operators monitoring several screens can be supported by systems that track gaze behavior and automatically notify the operator to adjust their scanning pattern \cite{Marois2020}. In a simulated task, Veerabhadrappa, et al. \cite{Veerabhadrappa2021} found that participants achieved higher performance on a simulated UAV refuelling task when they maintained a longer gaze on the relevant region of interest compared to less relevant regions. Further, in training scenarios, gaze-based measures can identify operator attention allocation and quantify progress of novice operators \cite{Shahab2021}. Gaze-based measures of novices can also be compared with those of experts to determine training progress and ensure efficient use of gaze.

In a review paper focused on pilot gaze behaviour, Ziv \cite{Ziv2016} found that expert pilots maintain a more balanced visual scanning. Expert pilots scan the environment more efficiently and spend less time on each instrument compared to novices. However, in complex situations, experts spend more time on the relevant instruments which enables them to make better decisions than novices. Overall, Ziv concluded that the differences in gaze behavior between expert and novice pilots are related to differences in flight performance.

\subsection{On object detection}


Object detection identifies various objects in the image such as cars, planes, dogs and cats and localize them, often using a bounding box around each object instance.

A plethora of good models have been developed for object detection \cite{zaidi2022survey}. They can mainly be classified into two categories, two-stage object detectors, and single-stage object detectors. With two-stage detectors, such as R-CNN \cite{girshick2014rich}, SPP-net \cite{He2015} and DetectoRS \cite{DetectoRS2020}, the first step involves creating region proposals, and the second step further refines the proposed bounding boxes and classify them. Whereas single-stage detectors directly generate bounding boxes and classes without the need for region proposals. Examples of those are YOLO \cite{YOLOv72022}, SSD \cite{liu2016ssd}, and EfficientDet\cite{tan2020efficientdet}. Generally speaking, two-stage detectors are more accurate than single-stage detectors but less applicable in real-time scenarios.

While these solutions are common for robotic perception \cite{Javier2018} and static sensing \cite{zou2019object}, only a handful of works assess their potential to understand user behavior.

Geert Brône, et al\cite{Brone2011} combined gaze data with object detection algorithms to argue the effectiveness of using detection algorithms for automatic gaze data analysis. They also designed a proof of concept for their method but did not report the results. Another research \cite{callemein2018automated} investigated the use of two detection models (YOLOv2 and OpenPose) to relate the gaze location specifically to the interlocutor's visible head and hands during humans face-to-face communication. 

A more recent work \cite{stubbemann2021neural} trained a classification model on synthetic data to classify images cropped around the gaze location as a method to annotate volume of interests. The issue with this approach is that an image (or image crop) will be assigned only one class. Even if it contains more than one object it will be labeled as an image of the most prominent object, missing the cases where a person could be looking at multiple objects that are close to each other. Whereas object detection holds the potential to relay this information by identifying all objects in the image not just the most prominent one. Similarly, \cite{barz2021automatic} also used a classification model combined with image cropping, and compared it to using an object detection model. However, they only relied on available pretrained models without any fit to specific application.

Unlike those previous works, our work presents a model tailored and tuned to a realist application and use the tools to extract meaningful behavioral data with UAV pilots.

Closer to our research interest, Miller et al. \cite{l2021post} fine tuned a pretrained InceptionV2 detection model to annotate objects of interest in eye-tracking video data, nut also integrate it with body motion capture data. Their aim was to relate the gaze location with the participant's body position and other objects in the frame. Further, they deployed their method in a user study to investigate distance from gaze point to target object in a target interception task. Our work is different from theirs, in the application scenario, the tools we deploy (learning model and biophysics measures) and the fact that we provide a more in depth analysis of the user study.

Finally, object detection demands a lots of processing power: running state-of-the-art detection algorithms at 30 fps can hardly be done onboard, a wearable or even a more capable robot computer. Previous performance test were conducted between variants of YOLO in live situation for drone emergency landing \cite{nepal_comparing_2022} in terms of the mean average precision (mAP) and frames per second (FPS). Two model stands out : YOLOv3 for is speed and YOLOv5 for its accuracy.


\section{Operator biophysics}\label{sec:biop}

\begin{figure}[htp]
    \centering
    \includegraphics[width=.7\columnwidth]{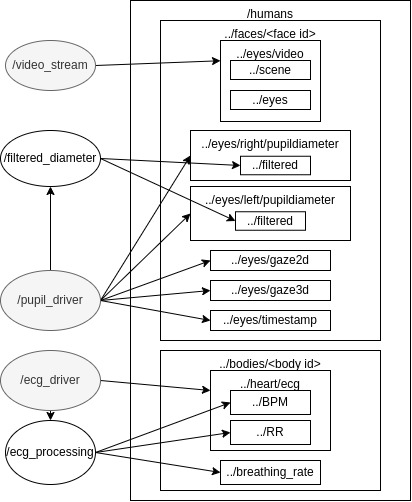}
    \caption{Our contribution to HRI4ROS pipeline shown as ROS rqt\_graph. The input nodes are on the left. Each node publishes in a set of specific topics under the standard namespaces.}
    \label{fig:rosgraph tobii and polar}
\end{figure}

Physiological signals capture and analysis are fundamental to human-centered robotic system design and validation. Several works demonstrate the relevance of these measurements, but each with its own implementation. It can then be time-consuming to reproduce any of these studies accurately and to deploy their tools and methods in other contexts.

In robotics, the Robot Operating System (ROS) positions itself as a standard to share and collaborate on code and software infrastructure. Its community is slowly integrating user-based measurements and wearable device drivers to cope with the reproductibility and sharing challenge of the Human-Robot Interaction (HRI) community~\cite{Mohamed2021}. A standard to include HRI considerations in ROS is currently in progress~\cite{ros4hriREP}.

Our work aims to contribute to the community effort with the integration of new sensor modality in the standard. The standard recommends to split the human-related aspects into five sub-namespaces. However, to better fit pupilometry data in the structure, we split the sub-namespace \emph{/human/faces/$<$face id$>$} in an other subcategory, namely the \emph{eyes}. Eyes data are related to gaze, pupil diameter and blinks. The gaze messages (2D and 3D) are compose of a \emph{geometry\_msgs/Point} and two timestamps : one from ROS, the other from the device (if available). The pupil messages (raw and filtered) publishes the diameter value in millimeters as a \emph{float64} alongside the same two timestamps. The pupil driver consider a web socket connection, generally over WiFi, compatible with several brands, namely Tobii and PupilLabs. Most of pupilometry commercial devices also grant access to live stream of the eyes camera and the scene camera, a frontal camera giving a first person view of the scene as viewed by the user. Another node subscribe to these streams and convert them in ROS image topics.

The second sensing modality is based on electrocardiogram signal. The driver provides the raw ECG values (in mV) in the \emph{/humans/bodies/$<$body id$>$} namespace. We also provide with more advanced features extracted live from the signal: RR intervals and beats per minutes. The breathing rate can also be extracted from the ECG signal, or, if available, from an acelerometer position on the chest. All these data are published as stamped \emph{float64} ROS messages.

Figure \ref{fig:rosgraph tobii and polar} illustrates our contribution to the ROS4HRI pipeline~\cite{ros4hriREP}, based on adding these two new sensing modalities. Each new user, or robot operator, is given a unique ID, based on the available modules to perform person identification. The current HRI4ROS pipeline~\cite{Mohamed2021} recommends using face recognition, but one could also leverage pupil shape as a unique biometric identifier using our contribution. The sensors reference frame can be link to the body with geometrical transformation (TF) messages. All these values are published live when using the devices, but they can easily be recorded in a ROSbag for post processing and data analysis. However, to monitor the data quality live and add some keypoints to the recordings, we also developed a Jupyter Notebook visualization interface. All code is publicly available online\footnote{https://git.initrobots.ca/aduval/bio\_physics}.

\noindent\textbf{Live data processing}

Most pupillometry devices provide gaze information in the eye camera frame as well as in the user space, following a calibration step using markers. However, pupil diameter and ECG cannot be used in their raw format to infer cognitive state.
A first pass on the pupil value prevents propagation of any physical outliers ($9<d(mm)<1$). Then the pupil diameter data is processed over sliding windows of 1 second duration with:
\begin{equation}
    d_{filtered} = d_{lp} - \tilde{d}_{base}
\end{equation}
where $d_{lp}$ is the raw value after a low pass filter was applied with cutoff frequency of 4Hz, and $\tilde{d}_{base}$ is the median of the baseline values. A baseline of at least 30 seconds must be recorded before running any experiment, for instance when the participant is undergoing the calibration process of the glasses.

As for the ECG data, we rely on a well maintained Python library popular among neuroscience academic projects, \emph{NeuroKit2}. We provide it with the raw signal, in sliding windows of 30 seconds, and it includes the signal cleaning step and feature extraction. However, since several features may be selected by the developer depending on the study context, our node provides only RR interval, BPM outputs and breathing rate extracted from the cleaned signal.

\section{Learn to see robots}\label{sec:learn}
\subsection{Dataset building}

Object detection is core to transferring the potential of gaze behavior analysis into UAV piloting task. Notwithstanding the accuracy of the existing models, it is known that a machine learning model is as good as the data used to train it. That dataset must contain enough images and instances for each class that the experiment targets. We collected our own dataset including four classes: UAV (drone), robot manipulator (arm), mobile robot base (rover) and remote controller. We selected classes that are available in our lab and that are more suitable to our experiment.

\begin{table}[h]
\centering
		{\caption{Dataset split details} \label{tab:DatasetDetails} } 
		
   \begin{tabular}{|c|c|c|c|}
   \hline
   
{\bf Class} &  {\bf No. Train. instances} & {\bf No. Valid. instances}  & {\bf All}    \\  \hline
drone       &   425     & 116  &    541 \\  \hline
robot arm  & 165       & 75    &    240 \\  \hline
mobile robot & 272      & 100   &   372\\  \hline
controller   & 217      & 77    &    294\\  \hline
Total       & 1079      & 368  &    1447   \\  \hline
		\end{tabular}
\end{table}

\begin{table*}[t]
\centering
		{\caption{Algorithms comparison results} \label{tab:AlgorithmsComparisonResults} } 
		
   \begin{tabular}{|c|c|c|c|c|c|}
   \hline
   
 {\bf Class}           &    {\bf V3T} &  {\bf V3D} &  {\bf V3SPP} &  {\bf V5} &  {\bf V7} 			\\  \hline
                 all   &    0.938 (0.958)&  {\bf 0.977} (0.981)&  0.971 {\bf (0.984)}&  0.974 (0.973)&  0.976 (0.982) \\  \hline
               drone   &    0.915 (0.935)&  {\bf 0.968} (0.969)&  0.954 (0.966)&  0.964 {\bf (0.974)}&  0.964 (0.968) \\  \hline
           robot arm   &    0.966 (0.991)&  {\bf 0.996} (0.995)&  0.98 (0.994)&  0.99 (0.995)&  0.994 {\bf (0.996)}	\\  \hline
        mobile robot    &    0.989 (0.995)&  0.992 (0.995)&  0.99 (0.995)&  {\bf 0.998} (0.995)&  0.997 {\bf (0.996)}	\\  \hline
          controller   &    0.879 (0.911)&  0.953 (0.964)&  {\bf 0.958 (0.98)}&  0.945 (0.929)&  0.947 (0.966)	\\  \hline

		\end{tabular}
  \vspace{-2em}
\end{table*}

For the UAV, we collected images of a small indoor collision-resilient quadcopter designed by de Azambuja et al.~\cite{cognifly2022}. For the robot arm, we picked several photos of the Gen3 lite robot arm from Kinova\footnote{\url{https://www.kinovarobotics.com/fr/produit/robot-gen3-lite}}. For the mobile robot, we used images of the Clearpath Dingo-D\footnote{\url{}https://clearpathrobotics.com/dingo-indoor-mobile-robot/}. Lastly for the controller, we selectted images of the PS4 Dualshock controller\footnote{\url{https://www.playstation.com/en-us/accessories/dualshock-4-wireless-controller/}}, one of the most common controller in mobile robotics and the one provided with the Clearpath base.

A total of 909 images with a total of 1447 object instances. To achieve this number and diversify the characteristics of the images, we leveraged four methods to collect those images. The first was to look into our lab's archive for videos and photos containing the target objects. Second, using a Python script, we automated download of relevant images from a search engine. Third, Downloading online public videos showing the objects and then extract frames from them. Finally, we took several new pictures and videos on our own, mostly to cope for the lack of public material on the custom UAV. We then annotate the images using an open source tool called LabelImg \cite{Tzutalin2015}.  

Furthermore, it is known that collecting and annotating images is a time consuming task and deep learning models needs a lot of data to train. So, we leveraged transfer learning to utilize pretrained models on COCO dataset \cite{Lin2014} then used our dataset for finetuing and validation.

For the dataset split, since the resulting dataset is  still rather small, we did not directly split the images to training and testing. Instead we tried to split them to maintain a percentage of around 80\% of each class instances for training and around 20\% for validation (some images may have 2 or 3 instances of an object and others may have 1 or 0). This resulted in 684 images for training and 225 images for validation. Table \ref{tab:DatasetDetails} represents the dataset and the split details.

\subsection{Algorithms comparison}

As mentioned in sec.~\ref{sec:litrev}, YOLO is well known for its object detection speed and accuracy with several versions of this model to choose from YOLOv1 to YOLOv7. Aiming at live detection eventually, we restrict our model selection to compare the YOLO algorithms family.

We first started with the evaluating of YOLOv3, YOLOv5 and YOLOv7 performances in our specific detection problem. For that, we selected three versions of YOLOv3: Tiny (V3T), default (V3D) and SPP (V3SPP), as well as the large version of YOLOv5 and YOLO7 largest model (e6e). Table \ref{tab:AlgorithmsComparisonResults} presents the results of the evaluation for each model pretrained in MS COCO, finetuned and evaluated on our collected dataset. Data augmentation was employed by the hyperparameters of YOLO during the models training to increase the effective size of the dataset and enhance the model generalization. The training and validation was conducted on Compute Canada clusters \cite{baldwin2012compute}. For each model, we reported F1 measure and mean average precision (between practices) at intersection over union of 0.5.

Secondly, we needed to be able to use those trained detection models within ROS to fit into the pipeline detailed in sec.~\ref{sec:biop} and to align the detection information with the person's gaze information. To do that, we decided to look into the publicly available open source ROS packages for YOLOv3 \cite{bjelonicYolo2018}, YOLOv5 \cite{vasilopoulosYolov3ros_2019} and YOLOv7 \cite{lukazsoyolov7ros2021} using the weights from our training. When running the feed video through each model in live, we end up getting 8 FPS, 3 FPS and 1 FPS for YOLOv3, YOLOv5 and YOLOv7 respectively. The tests were conducted on an Intel® Xeon(R) E-2276M CPU @ 2.80GHz × 12 CPU, with a NVIDIA Quadro P620 512 CUDA core GPU with 4 GB GDDR5, 16 GB DDR4 RAM running Ubuntu 20.04.4 LTS and ROS Noetic.

\subsection{Gaze to objects}

\begin{figure}[htp]
    \centering
    \includegraphics[width=0.7\columnwidth]{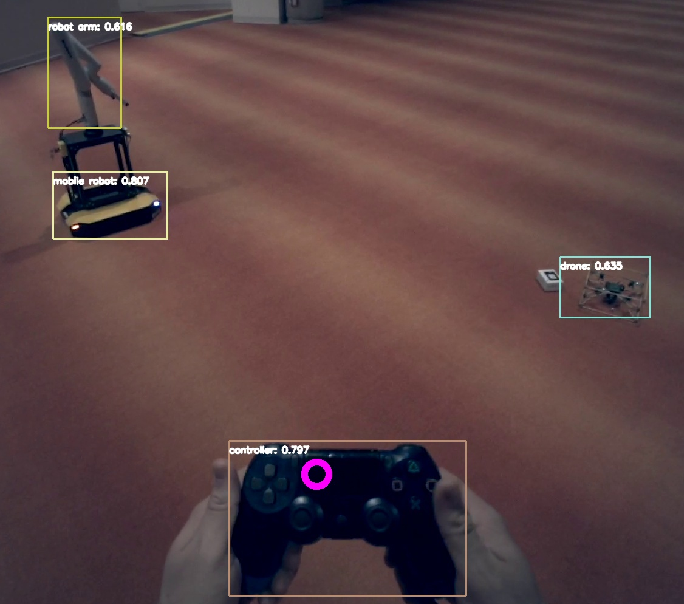}
    \caption{Gaze on a dualshock 4 controller. A person was asked to look for 10 seconds at the remote controller. In pink we see the gaze. Each object in the video frame is bound and tagged with the class and confidence of prediction.}
    \label{fig:gaze on object}
    \vspace{-1em}
\end{figure}

Gaze tracking in a 3D environment is easy by itself. But combining gaze tracking on object in a dynamic 3D space makes it more difficult. Detected objects from the video feed have bounding boxes around them. Adding the gaze coordinates base of the same image reference, we can deduct what the person is looking at. Like shown in Figure \ref{fig:gaze on object} the gaze is inside a bounding box, it means the person is looking on that particular object, in this case, a PS4 controller.

\begin{figure}[htp]
    \centering
    \includegraphics[width=0.45\columnwidth]{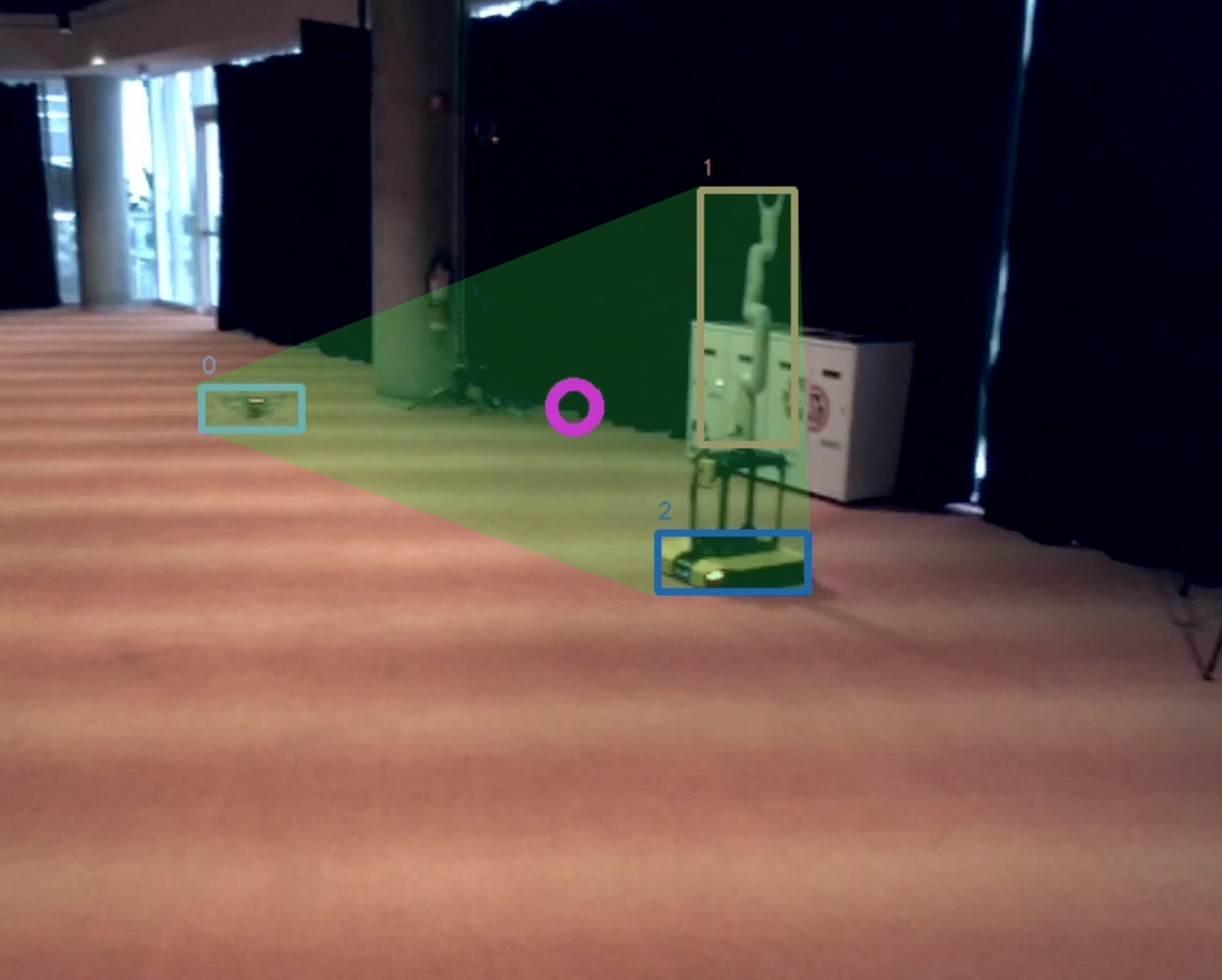}
    \includegraphics[width=0.45\columnwidth]{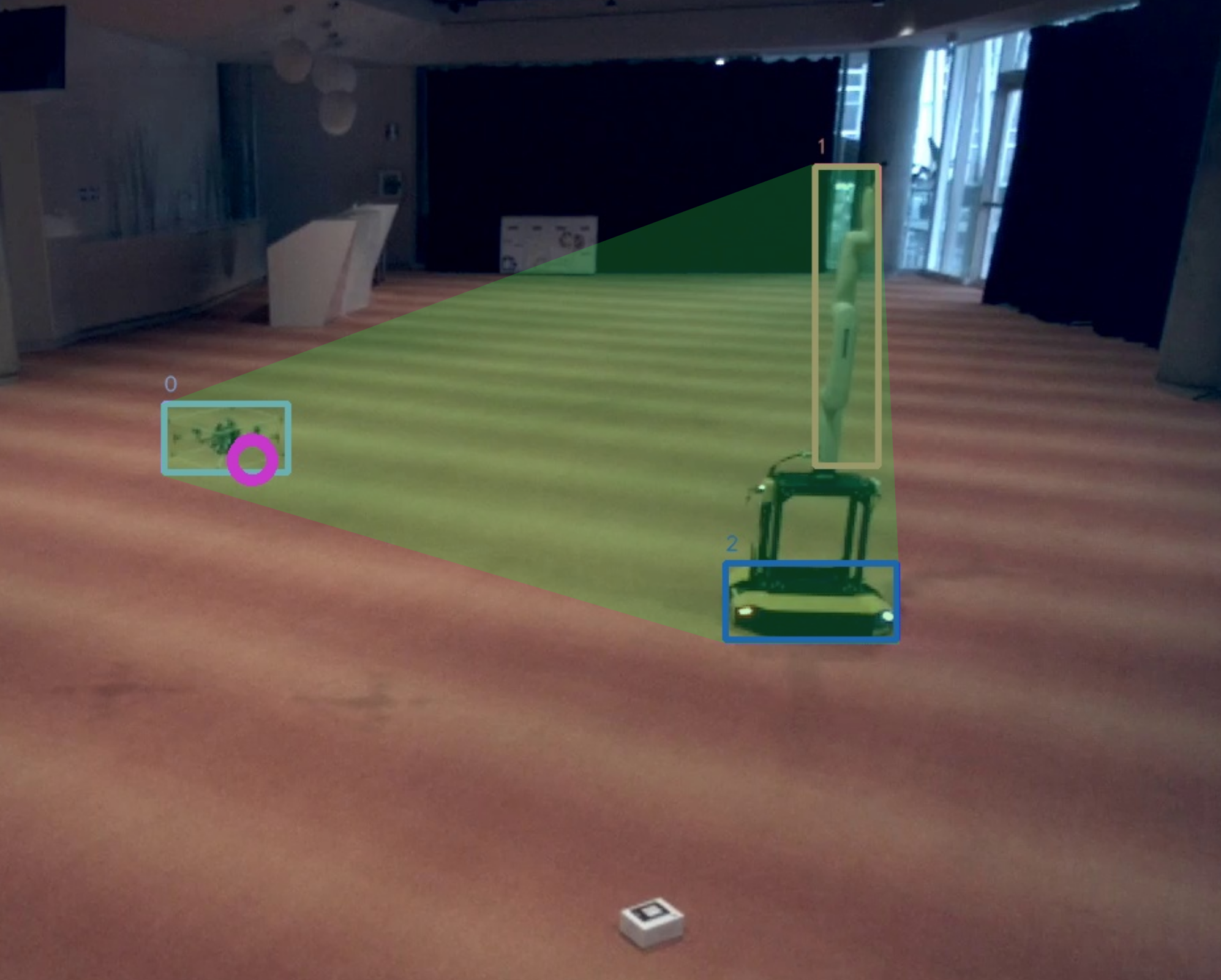}
    \caption{The gaze position within the ROI (green). The ROI is dynamic to follow the moving objects. The gaze (pink) is moving in this area between the drone and the robot arm.}
    \label{fig:gaze roi}
    \vspace{-1.5em}
\end{figure}

To avoid any collision, the person needs to sweep the gaze over the surrounding objects. To note if the person is looking at the action or not, we added a dynamic ROI (Region Of Interest) that stay attached to the detected object in the scene. We can see the ROI in green in the Figure \ref{fig:gaze roi}. We use functions made by Dinu C. Gherman to define the ROI by giving points in an image\footnote{\url{https://www.oreilly.com/library/view/python-cookbook/0596001673/ch17s19.html}}.



\section{User study}\label{sec:exp}
\subsection{Method}

\begin{figure}[htp]
    \centering
    \includegraphics[width=8cm]{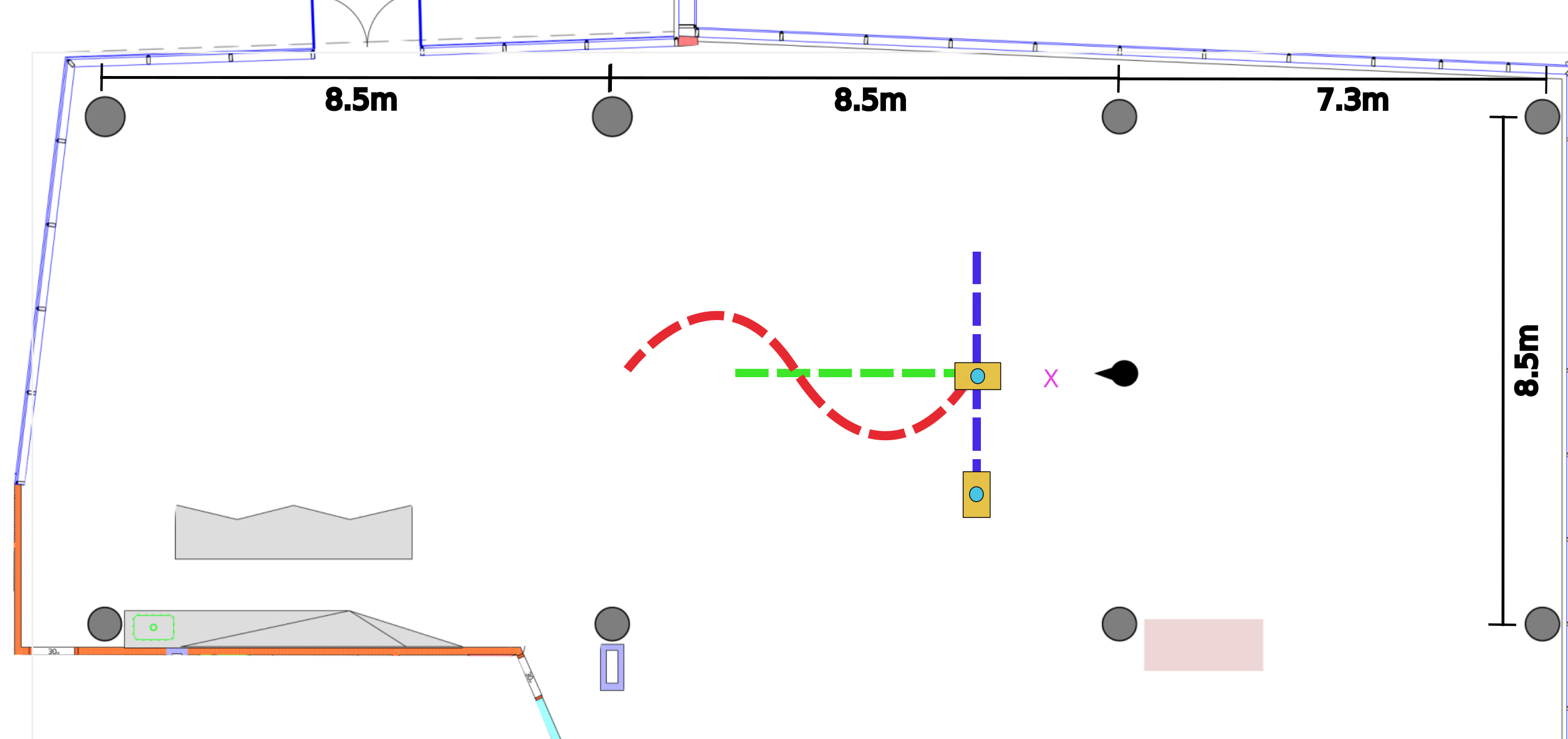}
    \caption{Map of the indoor environment. The black dot represent the participant. The yellow rectangles with blue dot are the Dingo with the Kinova arm on top on their starting point. Blue line is the side to side path. Green line the back and forth path. Red curve the slalom path. The pink X represent the Cognifly drone.}
    \label{fig:map}
    \vspace{-1.5em}
\end{figure}

The study was conducted indoors to maintain a more controlled environment. Figure \ref{fig:map} shows the map of the environment. A large and long room with a ceiling height of 3 meters. Participants (n = 8) piloted a drone around a stationary arm on top of a robot (either stationary or moving, see conditions below and Fig.~\ref{fig:gaze roi}) with the goal of flying the drone around the arm as many times as possible. Biophysical measurements, including pupil diameter, eye gaze, and heart rate (ECG), were recorded from each participant during each of the tasks. The study was divided into 5 tasks, with a duration of 1 minute each:

\begin{enumerate}
  \item Pilot the drone freely with no obstacles around (baseline condition to familiarize participants);
  \item Pilot the drone around a static vertical Kinova robot arm affixed to a Clearpath Dingo-D;
  \item Pilot the drone around a vertical Kinova arm affixed to a Clearpath Dingo-D moving back and forth at 25 cm/s on a distance of 4 meters;
  \item Piloting the drone around a vertical Kinova arm affixed to a Clearpath Dingo-D moving side to side at 25 cm/s on a distance of 4 meters;
  \item Piloting the drone around a vertical Kinova arm affixed to a Clearpath Dingo-D moving in a slalom motion .
\end{enumerate}

Prior to the experiment, each user was asked about their experience with drone piloting. This needed to be considered because it can have a great impact on their performance. A person piloting a drone for the first time will not have the same reflexes or familiarity as someone with many hours of flight.

Before starting the tasks, the participant put on the Polar H10 heart rate chest band sensor under the shirt to record heart rate measures and a pair of Tobii Pro Glasses 3 eye-tracking glasses. Then, we explained the controls of the Cognifly drone~\cite{cognifly2022}. Participants were instructed to stand still on a mark on the ground (black dot in Figure \ref{fig:map}).\

To establish a baseline of luminance for each of the areas participants would be looking at, we had the participant look for 10 seconds each at the remote controller held in their hands, the drone, the Dingo-D, and the Kinova robot arm. Once done, we began Task 1.

\begin{figure}[htp]
    \centering
    \includegraphics[width=.75\columnwidth]{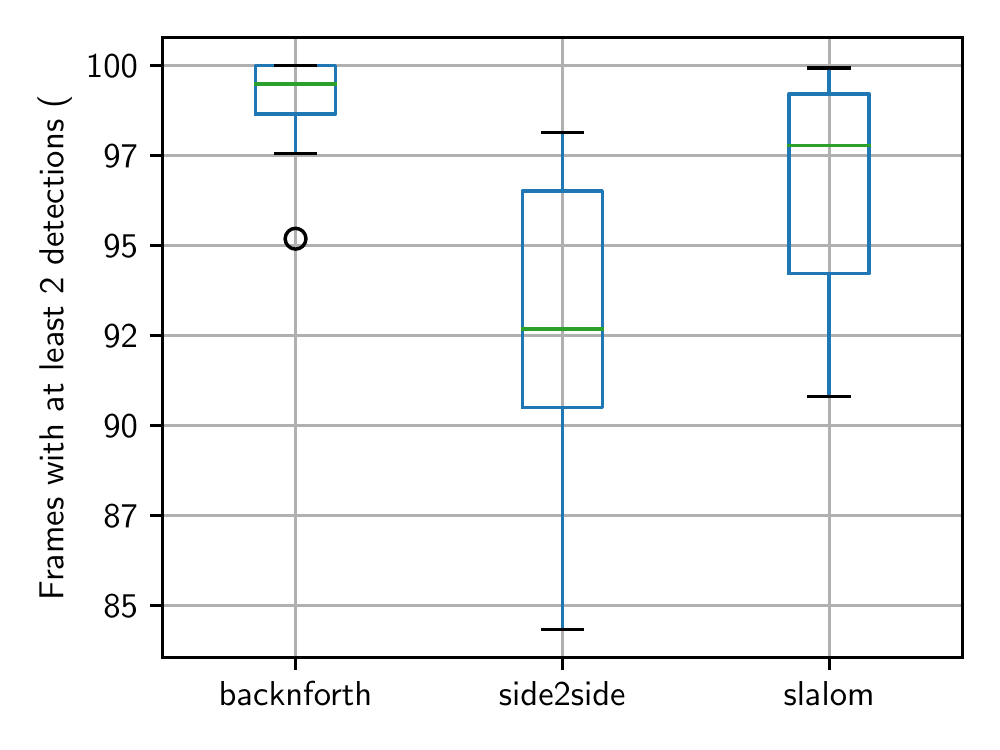}
    \caption{Detector performance: percentage of frames with at least 2 objects detected out of 3 (arm, base, uav).}
    \label{fig:classperf}
    \vspace{-1em}
\end{figure}

After completing Task 1 (freely flying the drone in the area), participants completed Task 2 (stationary condition). After Task 2, participants completed either Task 3 (back and forth motion) or 4 (side to side motion). Tasks 3 and 4 were counterbalanced, so all participants did both, but not in the same order. Finally, participants completed Task 5 (slalom motion). During the experiment, drone crashes occurred for some participants (n=3). There is no particular task that had more crashes than others. In these cases, the task was restarted.
Figure~\ref{fig:classperf} show the classifier distribution over all participant for the last three tasks: almost every frame had at least two objects detected, enough to compute a meaningful ROI.

\subsection{Results}

\begin{figure}[htp]
    \centering
    \includegraphics[width=.7\columnwidth]{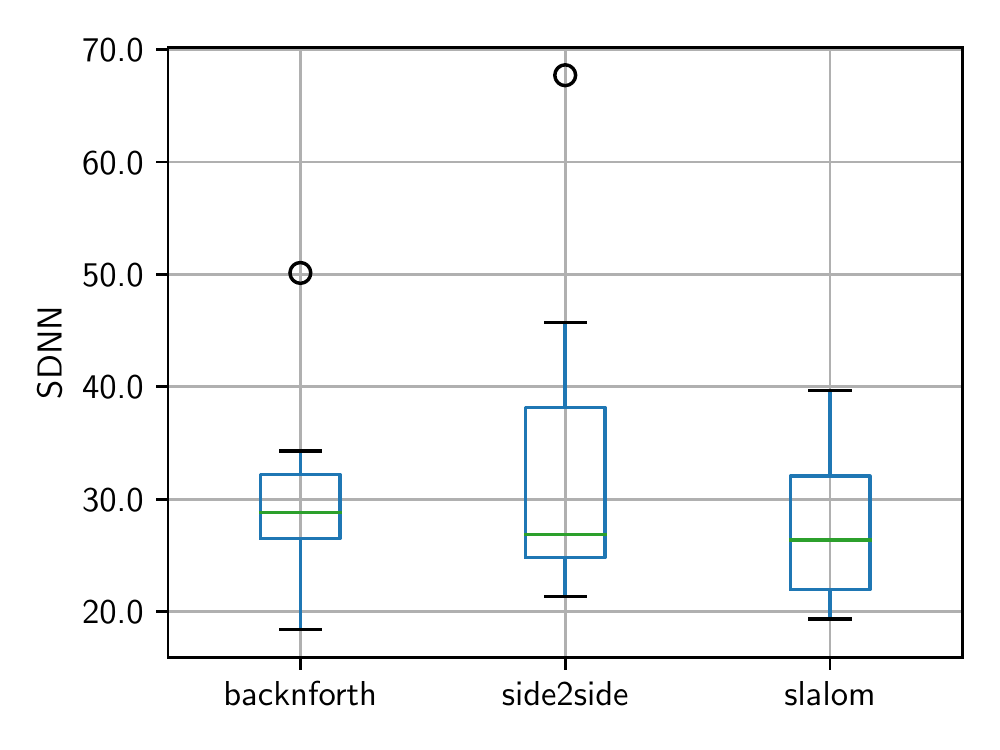}
    \includegraphics[width=.7\columnwidth]{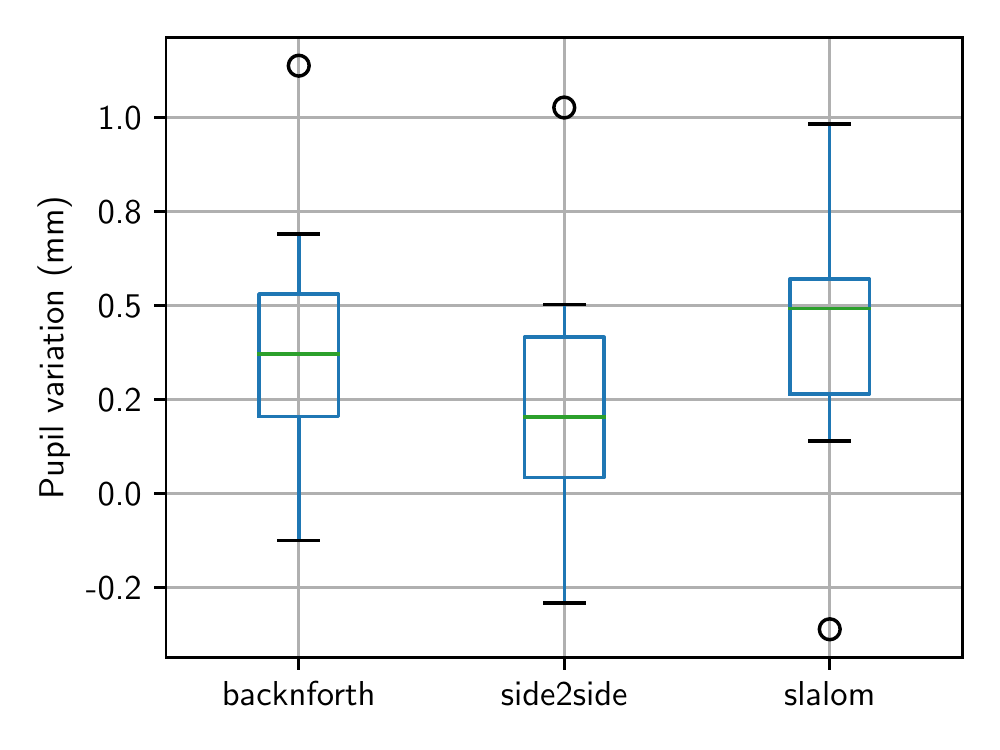}
    \caption{Cognitive load metrics average for all participants by task.}
    \label{fig:CLtask}
    \vspace{-1em}
\end{figure}

On average, participants completed the most circles around the Kinova arm in the back and forth (Task 3) and side to side (Task 4) conditions, and the least in the slalom condition (Task 5), suggesting that the slalom condition was most challenging. Pupil diameter measures support this idea, as participants had the largest pupil diameter in the slalom condition, on average (see Figure \ref{fig:CLtask}). Heart rate variability measures (SDNN) showed a small trend towards this pattern as well between the back and forth and slalom tasks.

\begin{figure}[htp]
    \centering
    \includegraphics[width=0.7\columnwidth]{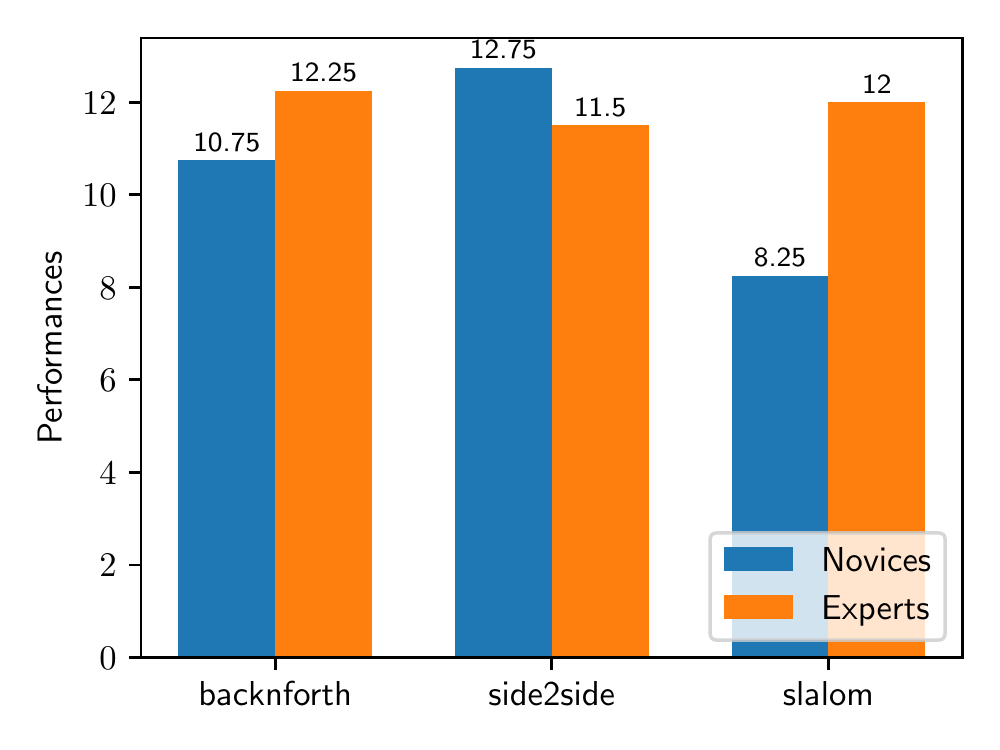}
    \caption{Average number of circles around the obstacle (arm) by group for the final three tasks. All participants performed the slalom task last. Most novices performed the back and forth task before the side to side task, and the reverse for most experts.}
    \label{fig:performance}
    \vspace{-1em}
\end{figure}

We then split the data between expert and novice drone pilots, with experts having at least 15 hours of flight experience. There were four experts and four novices with this division. For overall performance, experts outperformed novices in all tasks except the side to side task (see Figure \ref{fig:performance}). However, this could be due to additional task-specific training. Due to the counterbalancing (which did not take expertise into account), three out of the four novices performed the side to side task later in the experiment than most of the experts. The expert group maintained similar performance levels across tasks, while the novice group had a drop in performance for the slalom condition.

\begin{figure}[htp]
    \centering
    \includegraphics[width=.7\columnwidth]{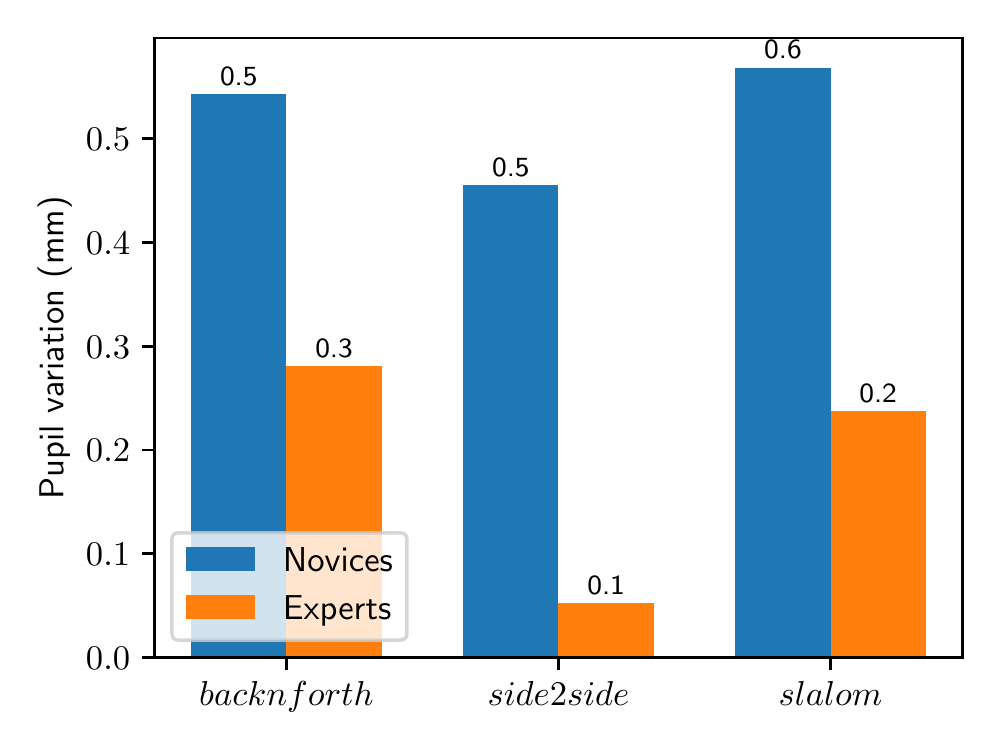}
    \includegraphics[width=.7\columnwidth]{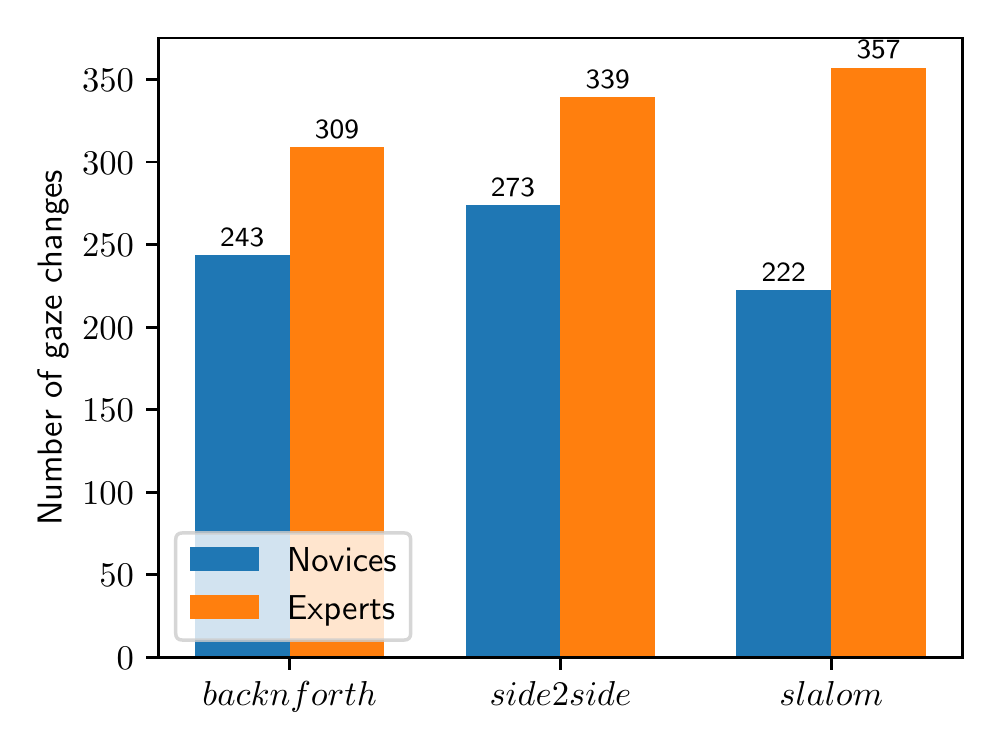}
    \caption{Differences in pupil diameter variance (top) and gaze changes (bottom) between experts and novices.}
    \label{fig:CLtaskgroups}
    \vspace{-1.5em}
\end{figure}

The pupil measures differed slightly between the groups (see Figure \ref{fig:CLtaskgroups}), with expert pilots having the largest pupil diameter in the slalom condition, followed by the back and forth condition, and finally the side to side condition. Novice pilots had the largest pupil diameter in the back and forth condition, followed by slalom, and then the side to side condition. This suggests that cognitive load in expert pilots may have been highest in the slalom condition, whereas in novice pilots, cognitive load may have been highest in the back and forth condition.

For the heart rate variability measures, experts had higher SDNN for the back and forth condition than for the side to side or slalom conditions. Novices had higher SDNN for the side to side condition, less for the slalom condition, and least for the back and forth condition. As with the pupil and performance measures, this pattern of results shows a trend of familiarization, as each group experienced increased SDNN (an indicator of lower cognitive load) in the condition most in the group performed fourth. 

\begin{figure}[htp]
    \centering
    \includegraphics[width=4cm]{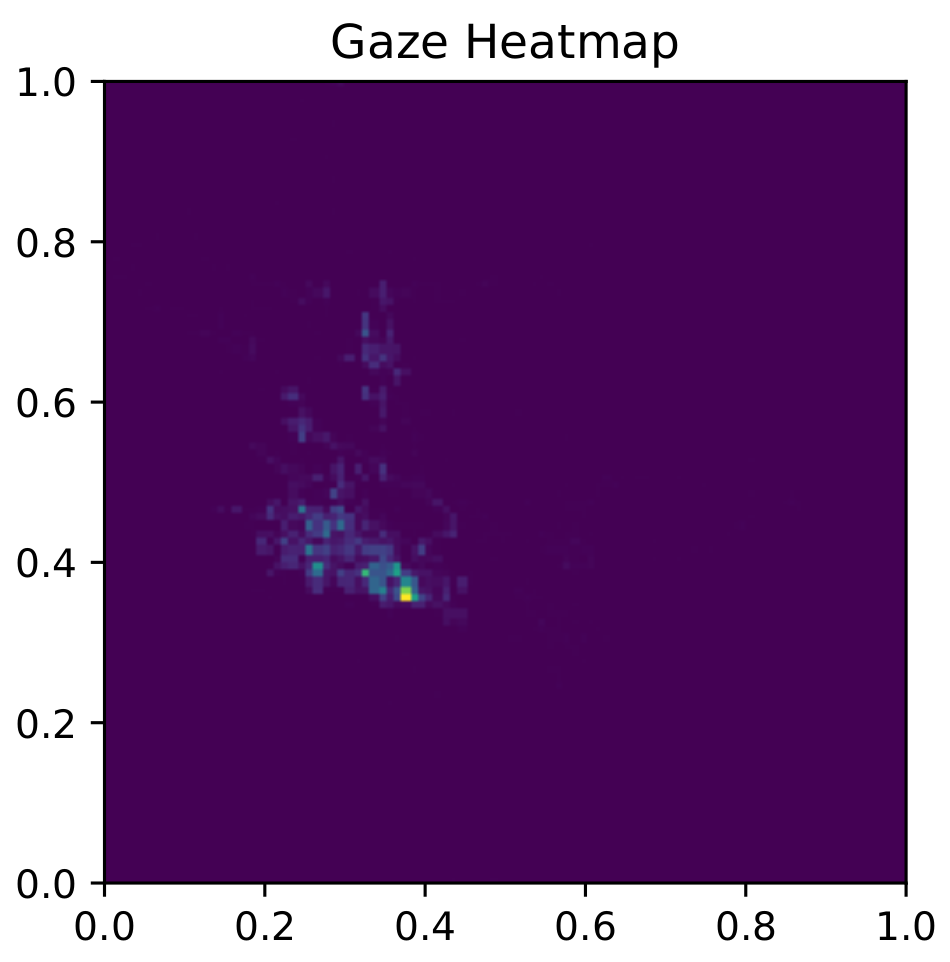}
    \includegraphics[width=4cm]{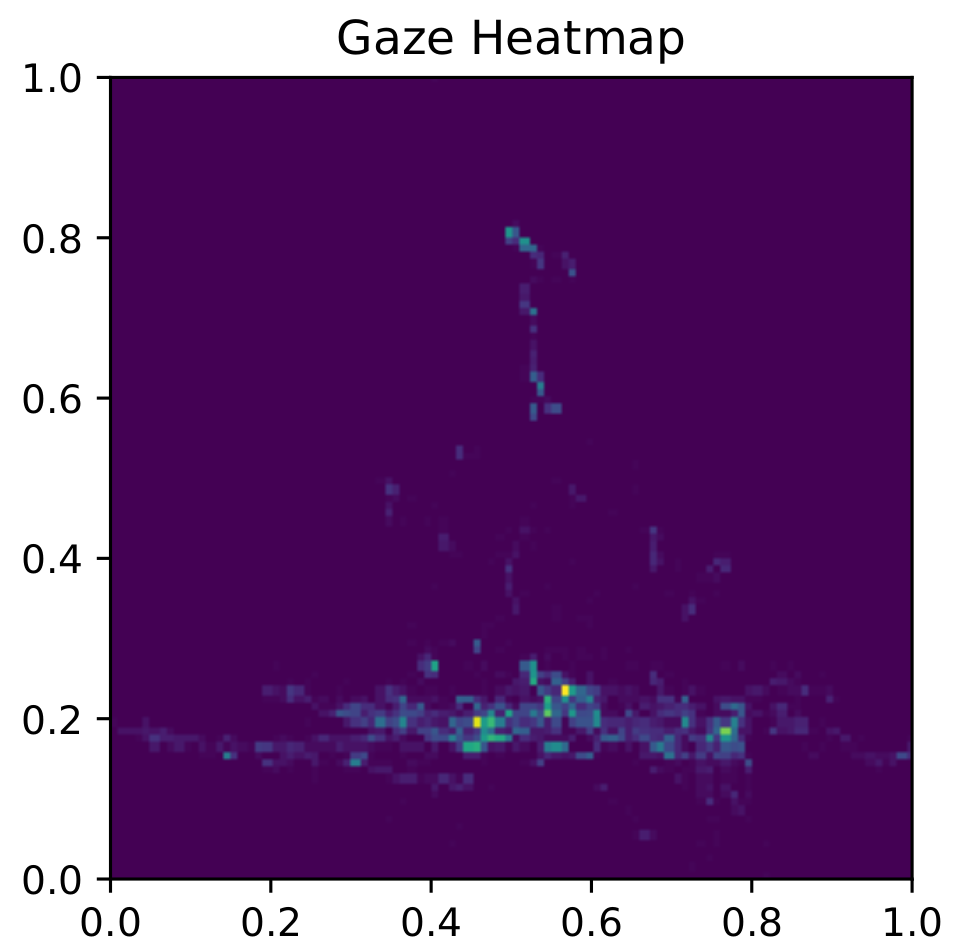}
    \caption{The gaze heatmap (gaze spatial distribution) for the slalom task: left for an expert operator, and right, for a novice.}
    \label{fig:gazehm}
    \vspace{-1em}
\end{figure}

Finally, as seen in Figure \ref{fig:CLtaskgroups}, experts shifted their gaze more than novices, but their gaze was more focused and intentional (see heatmap in Figure \ref{fig:gazehm}). This may suggest that experts were more comfortable with piloting a drone and were able to monitor the area more efficiently to perform the tasks compared to novices.

\section{Conclusion}
In order to better understand how UAV pilots operate, we introduced a complete integration of pupil and heart rate features into the HRI4ROS pipeline. We presented a set of ROS nodes to capture and process live data from these sensors and we built a dataset to train a model for object detection on pilot frontal camera feed. Our trained algorithm showed detection performance close to perfect. We then conducted a short user study, showing gaze behaviors that differ between expert and novice pilots and comparable cognitive load variation between tasks of different difficulty. Ultimately, we will extend this study to more participants and explore the relation with flight patterns: experts seem to use yaw control more than novice.


\bibliographystyle{IEEEtran}
\bibliography{main}

\end{document}